\documentclass[12pt,twoside,english]{article}

\usepackage[T1]{fontenc}
\usepackage[latin9]{inputenc}
\usepackage[a4paper]{geometry}
\usepackage{amsmath}
\usepackage{graphicx}

\makeatletter

\providecommand{\tabularnewline}{\\}

\newcommand{\lyxaddress}[1]{
\par {\raggedright #1
\vspace{1.4em}
\noindent\par}
}

\@ifundefined{showcaptionsetup}{}{%
 \PassOptionsToPackage{caption=false}{subfig}}
\usepackage{subfig}
\makeatother

\usepackage{babel}
\begin{document}

\title{Seasonal trend of AOD and \AA ngstr\"om exponent($\alpha)$ over
Indian megacities in varying spatial resolution.}

\author{Subhasis Banerjee, Sanjay K. Ghosh}

\maketitle

\lyxaddress{Center for Astroparticle Physics and Space Science, Bose Institute, Kolkata}
\begin{abstract}
Aerosol optical characteristics AOD and  \AA ngstr\"om exponent
is often used to asses environmental aerosol loading. AOD or Aerosol
Optical Depth is an indirect measure of atmospheric aerosol loading
by means of total extinction of incoming solar radiation due to scattering
and absorption whereas  \AA ngstr\"om exponent($\alpha)$ is used
to get qualitative understanding of aerosol particle size. Analysis
of long term time series AOD data reveals how AOD vis-à-vis aerosol
on a particular place changes over time. Similar study with  \AA
ngstr\"om exponent($\alpha)$ gives an idea how particle size distribution
is changing over some area. Such studies cannot be conducted by data
measured by ground based stations alone because they are inadequate
in numbers on earth moreover such data for considerably long period
are not available for most places. To overcome this, radiance data
sensed by MODIS instruments on board Aqua and Terra satellite have
been used by many authors. In this study 13 years of MODIS level 2
data by Terra have been analyzed for four Indian mega-cities namely
Delhi, Mumbai, Kolkata and Chennai using well known Mann-Kendall statistics
and Sen's non-parametric estimation of slopes.
\end{abstract}

\section{Introduction }

Aerosols or particulate matters influences earth's climate system
by absorbing and reflecting solar radiation\cite{kauffman}. Aerosol
also modifies cloud properties. Several studies have shown detrimental
effects of aerosol on human health\cite{Molina,pope-1,pope-2,samoli}.
But it's impacts on climate and human health has not yet been completely
understood. Aerosols are released into the atmosphere by anthropogenic
and natural sources and even formed in the atmosphere by natural processes\cite{seinfeld}.
Considering the impact of aerosol on climate system and human health
several attempts have been made worldwide to monitor it's level. But
surface level monitoring stations are very rare. Satellite based measurement
have been increasingly in use to complement ground based measurements.
Satellites measures reflected solar radiation at the top of the atmosphere.
Column integrated value of aerosol are routinely estimated from those
measurements using suitable inversion algorithms\cite{remer-1}. Earth
Observing System satellites Aqua and Terra take daily nearly global
measurements in wide spectral range. Some studies have already used
those data to estimate long term trend of aerosols over megacities
worldwide. In such studies AOD data over a larger area around cities
have been used. In most studies, MODIS level 3 data have been used\cite{pinhas,gupta}.
Pinhas et al \cite{pinhas}studied AOD trend over megacities using
level 3 MODIS data along with MISR but they have not considered seasonal
effects. Level 3 data are provided in only 1 x 1 degree square grid
sizes . For consideration of smaller areas around cities level 2 data
must be used. However in their study they also found overall increasing
trend of AOD in Indian megacities other than Delhi. Here we have used
MODIS level 2 data in order to estimate long time trend of aerosol
loading over four Indian megacities in varying grid sizes considering
seasonal effects also. Gupta et al \cite{gupta}studied level 2 MODIS
AOD data over Karachi and Lahore in Pakistan. They also found high
aerosol loading near city centres compared to outskirts, as expected.
They reported considerable seasonal variation in their analysis. Acharaya
et al studied seasonal variability in AOD over India using MODIS level
2 land data\cite{acharaya}. In their study they captured increasing
trend as well as seasonal variability. Ramachandran et al studies
seasonal and annual mean trend of AOD from MODIS level 2 data over
different location in India\cite{rama}. Indian subcontinent has witnessed
unprecedented industrial activities in the last century. Population
in the cities has grown to manifold\cite{population}. Not surprisingly,
four mega-cities in India have shown tremendous increase in human
activities resulting tumultuous increase in atmospheric pollution.
Here, we have used well known Mann-Kendall statistics to estimate
seasonal trend of AOD and  \AA ngstr\"om exponent($\alpha)$ \cite{angstrom}
around the cities in different grid sizes. Sen's non parametric method
used to estimate true slopes.

\section{Data}

In our study we have used MODIS level 2 high resolution 10 km x 10
km (collection 5.1) aerosol retrieval (product id MOD04L2, Downloaded
from LAADSWeb website http://ladsweb.nascom.nasa.gov/). MODIS instrument
on board Terra and Aqua satellite takes almost near global measurements
in a wide spectral range (0.41-15 $\mu$m)\cite{remer-1}. AOD over
land and ocean are derived using these measurements. Different MODIS
algorithms are used to derive aerosol products over land and oceans.
In case of land retrieval dark-target approach is used, i.e., pixels
for which surface reflectance fall within 0.01 to 0.25 at 2.13 $\mu m$
are used\cite{remer-1}. Aerosol Optical Depth is retrieved for two
visible wavelengths of 0.47 $\mu m$ and 0.66 $\mu m$. AOD at 0.55
$\mu m$ actually interpolated using these values. In case of ocean
AOD is provided for seven wavelength (0.47 $\mu m$ to 2.13 $\mu m$).
In our study quality controlled\cite{remer-2} 'joint land and ocean'
product ``Optical\_Depth\_Land\_And\_Ocean'' (0.55 $\mu m$ )has
been used to study time trend of aerosol optical depth and ``Angstrom\_Exponent\_Land''
for  \AA ngstr\"om exponent (0.47 $\mu m$ to 0.67 $\mu m$)in four
Indian megacities. MODIS data have been collected over a area of 1
x 1 degree centered at each city of our interest and spatial averages
taken for 4 different grid sizes for further analysis. Four different
grid sizes of 0.25,0.50.0.75,1 degree squares have been taken to estimate
whether city center behaves differently respect to trends in aerosol
loading. In this paper all AOD data presented is at 0.55 $\mu m$
and all  \AA ngstr\"om exponent values are for 0.47 $\mu m$ to
0.67 $\mu m$.

\section{Methodology}

Mann-Kendall statistical test is a non-parametric test commonly used
to identify monotonic trend in meteorological time series data. In
this test Mann-Kendall statistic S is computed by\cite{gilbert,mann}
\begin{equation}
S=\underset{i=1}{\overset{n-1}{\sum}\,}\underset{j=i+1}{\overset{n}{\sum}}sgn(x_{j}-x_{k})
\end{equation}

where 

\begin{equation}
sgn(\theta)=\begin{cases}
\begin{array}{cc}
+1 & if\\
\,0 & if\\
-1 & if
\end{array} & \begin{array}{c}
\theta>0\\
\theta=0\\
\theta<0
\end{array}\end{cases}
\end{equation}

Here $x_j$ is the sequential data value and $n$ is the number of
data points. The null hypothesis $H_0$ is that the data come from
a population where random variables are identically distributed and
independent. The alternative hypothesis is that a monotonic trend
exists in the data series. The mean $E(S)$ and variance $\sigma^2$
of the statistic $S$ under $H_0$ are computed as: 
\begin{equation}
E(S)=0
\end{equation}

and 
\begin{equation}
\sigma^{2}=\frac{n(n-1)(2n+5)-\overset{q}{\underset{p=1}{\sum}}t_{p}(t_{p}-1)(2t_{p}+5)}{18}
\end{equation}

where $t_p$ is the number data in the $p$th group and $q$ is the
number of tied groups. Now one can compute the standard normal variate
$Z$ by 
\begin{equation}
Z=\begin{cases}
\begin{array}{ccc}
\frac{S-1}{\sigma} &  & if\\
0 &  & if\\
\frac{S+1}{\sigma} &  & if
\end{array} & \begin{array}{cc}
 & S>0\\
 & S=0\\
 & S<0
\end{array}\end{cases}\label{Z}
\end{equation}

in a two tailed test for trend the $H_0$ should be rejected if $|Z| > Z_{1-\alpha/2}$
where $\alpha$ is the significance level. A positive (negative) value
of $S$ indicates an upward (downward) trend. If trend is present
true slope may be estimated by a simple non-parametric procedure developed
by Sen\cite{sen}. True slope $Q$ of trend is computed as the median
of all slopes between all the data pairs, where $Q$ is given by
\begin{equation}
Q=Median[\frac{x_{i}^{'}-x_{i}}{i'-i}]\begin{array}{ccccc}
 & for & all & i<i^{'}\end{array}
\end{equation}
$x^{'}$ and $x$ being the data values at times $i^{'}$ and $i$
respectively. The advantage of Sen's method over linear regression
method is that this method is not greatly affected by outliers and
it can compute slope when there is missing data. Lower and upper limits
of confidence interval about the true slope may be esimated as $M_{1}$th
and $M_{2}$th largest of the $N^{'}$ ordered slopes\cite{gilbert}.
Where 
\begin{equation}
M_{1}=(N^{'}-C_{\alpha})/2\label{M1}
\end{equation}

and
\begin{equation}
M_{2}=(N^{'}+C_{\alpha})/2\label{M2}
\end{equation}

and $C_\alpha$ is given by 
\begin{equation}
C_{\alpha}=Z_{1-\alpha/2}\sigma
\end{equation}
 $\alpha$ being the chosen level.

Mann-Kendall test was extended to take seasonality into account by
Hirsch et al\cite{hirsch} which is popularly known as seasonal Mann-Kendall
test. Seasonal Mann-kendall test consists of computing Mann-Kendall
test statistic $S$ and its variance $\sigma^2$ separately for each
seasons and then combining the results. Let $S_i$ and $\sigma_i{^2}$
are MK test statistic and variance respectively for $i$th seasons.
\begin{equation}
S_{mk}=\underset{i=1}{\overset{m}{\sum}S_{i}}
\end{equation}
\begin{equation}
\sigma_{mk}^{2}=\underset{i=1}{\overset{m}{\sum}\sigma_{i}^{2}}
\end{equation}

Now using Eqn.(\ref{Z}) $Z$ statistic is computed and referred to
standard normal table when number of seasons and years are sufficiently
large\cite{gilbert}. However in this approach one assumes homogeneous
trend direction in different seasons. If trend directions are different
i.e., upward in one season and downward in other season then they
may cancel each other leading to erroneous test results. It is thus
useful to test homogeneity of trend direction in different seasons.
Let, 
\begin{equation}
\chi_{total}^{2}=\sum_{i=1}^{m}Z_{i}^{2}
\end{equation}
\begin{equation}
\chi_{trend}^{2}=m\bar{Z}^{2}
\end{equation}
\begin{equation}
\chi_{homog}^{2}=\chi_{total}^{2}-\chi_{trend}^{2}
\end{equation}

where 
\begin{equation}
Z_{i}=\frac{S_{i}}{\sigma_{i}}
\end{equation}
and 
\begin{equation}
\bar{Z}=\frac{1}{m}\sum_{i=1}^{m}Z_{i}
\end{equation}

The $\chi_{homog}^{2}$ values are then compared to a a table of chi-square
distribution of $m-1$ degrees of freedom to reject(accept) the null
hypothesis, $H_0$, of homogeneous seasonal trend. In case $\chi_{homog}^{2}$
exceeds a predetermined $\alpha$ critical value we reject $H_0$
and then seasonal MK test becomes meaningless. In that case MK test
should be carried out for individual seasons separately. But if $\chi_{homog}^{2}$
does not exceed the $\alpha$ critical value $\chi_{trend}^{2}$ if
referred to chi-square distribution with 1 df to test for common trend
throughout all the seasons.

\begin{center}
\begin{table}
\begin{centering}
\begin{tabular}{|c|c|}
\hline
Season & Month\tabularnewline
\hline
\hline
Winter & December - February\tabularnewline
\hline
Premonsoon & March - may\tabularnewline
\hline
Monsoon & June - September\tabularnewline
\hline
Postmonsoon & October - November\tabularnewline
\hline
\end{tabular}
\par\end{centering}

\begin{centering}
\caption{Monthwise prominent seasons in India}

\par\end{centering}

\end{table}

\par\end{center}

\section{Results and Discussion}

In our study we have averaged the data by four prominent seasons witnessed
in Indian subcontinent\cite{chattyopadhya}. To understand spatial
variation of AOD, we have taken four grid sizes i.e., 1 x 1 degree
square (OD or One Degree), 0.75 x 0.75 degree square (TQD or Third
Quarter Degree), 0.50 x 0.50 degree square (HD or Half Degree) and
0.25 x 0.25 degree square (QD or Quarter Degree).
\begin{figure}

\subfloat[Delhi]{\includegraphics[angle=-90,scale=0.3]{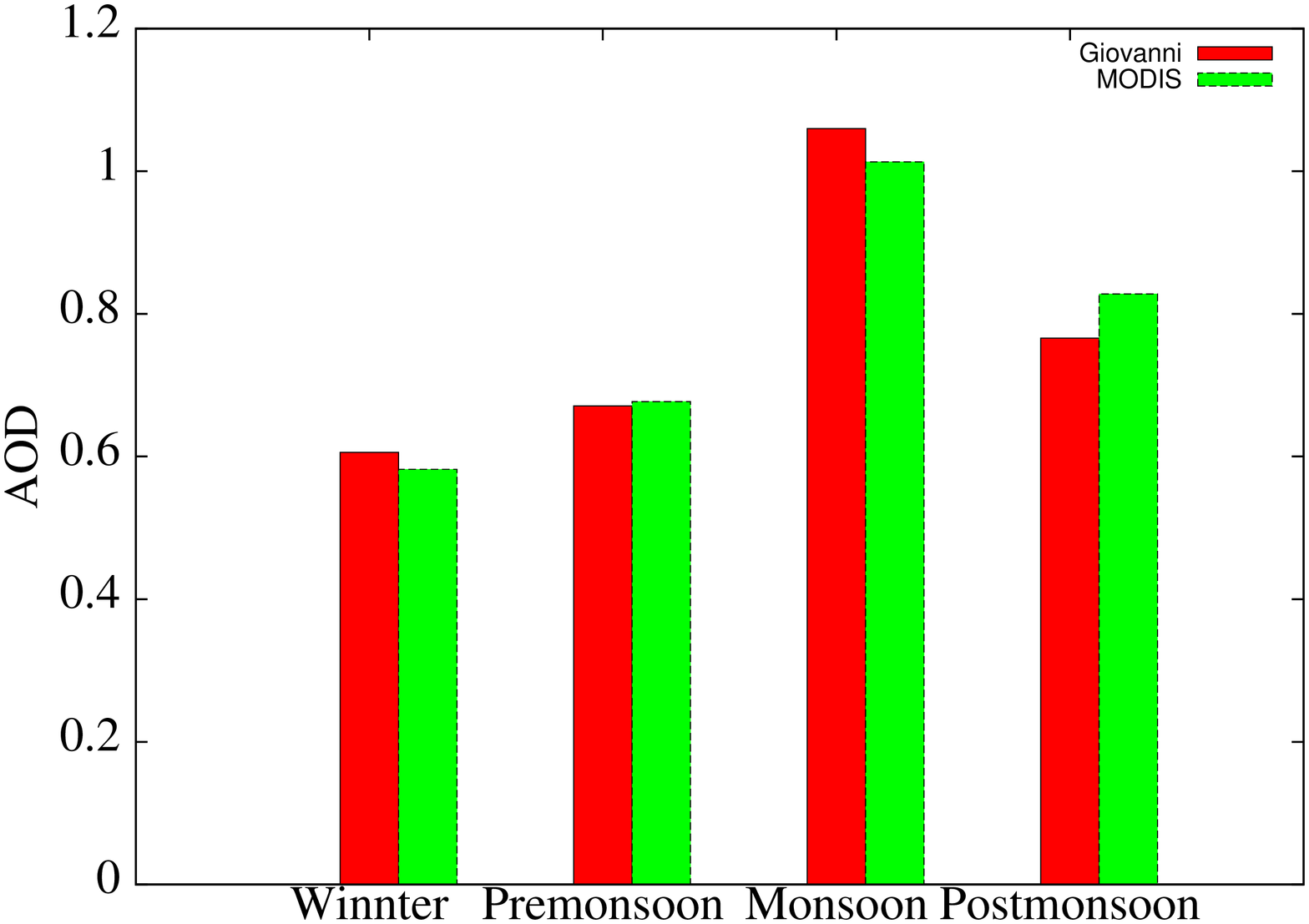}
}\subfloat[Mumbai]{\includegraphics[angle=-90,scale=0.3]{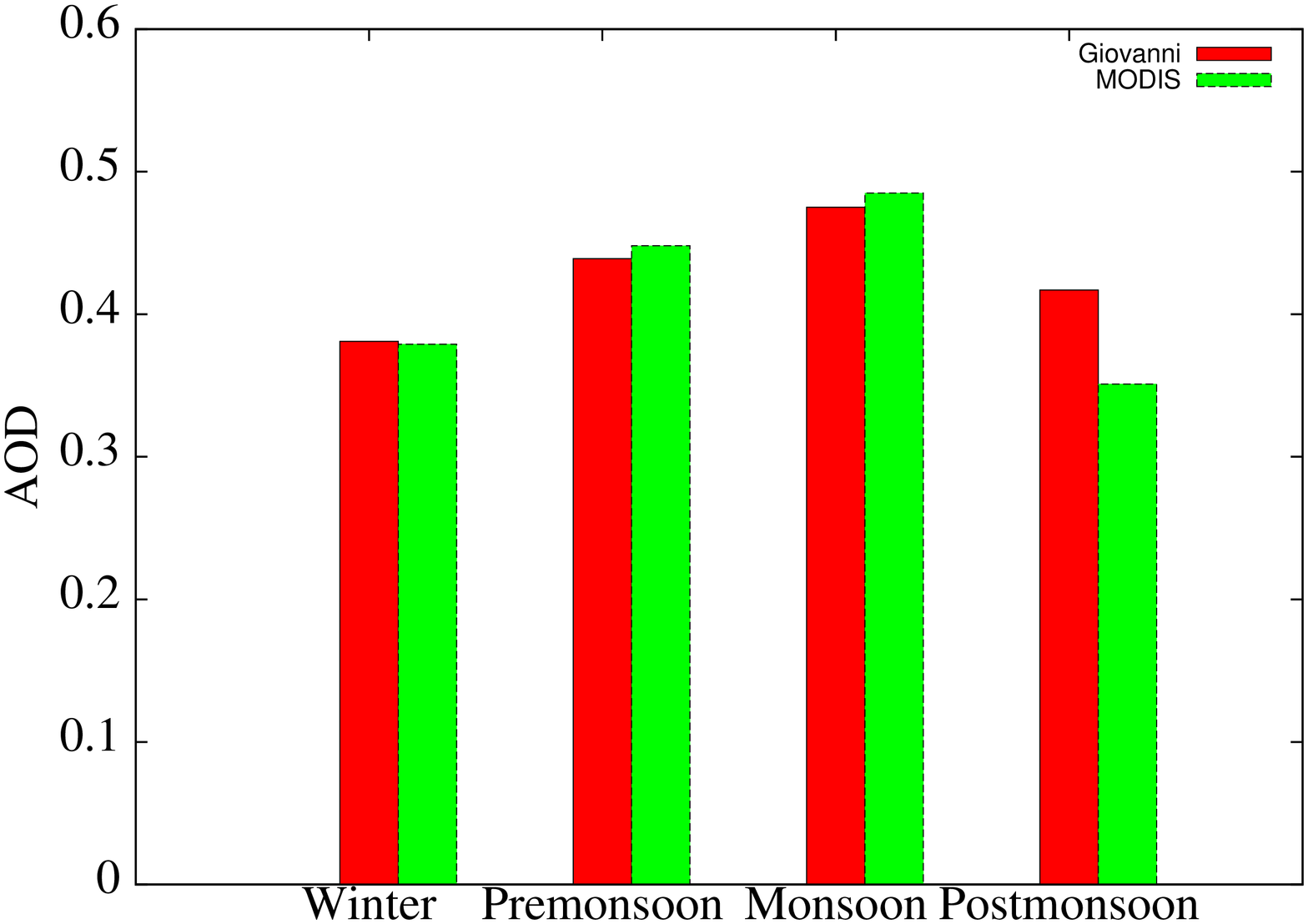}}

\subfloat[Kolkata]{\includegraphics[angle=-90,scale=0.3]{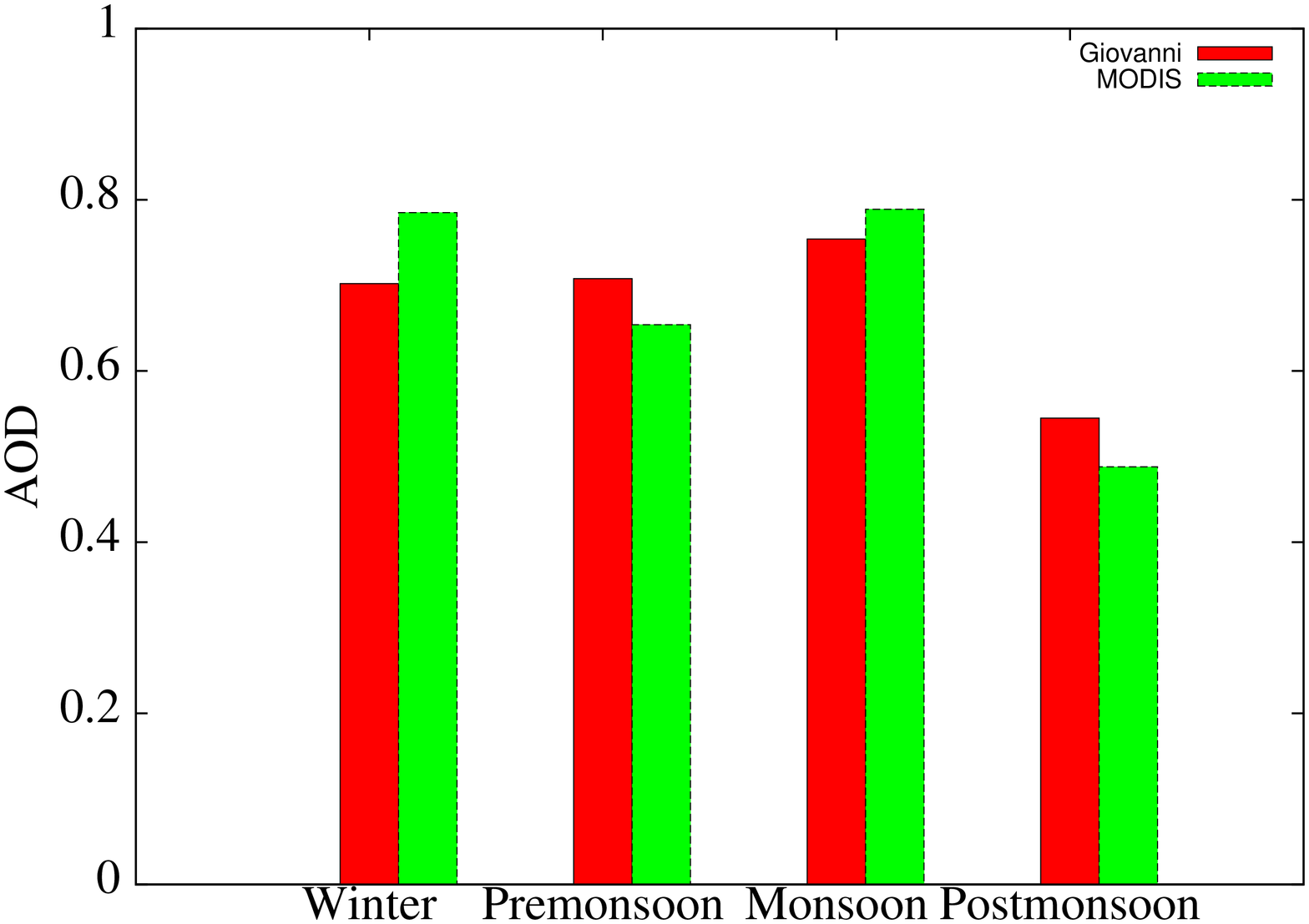}
}\subfloat[Chennai]{\includegraphics[angle=-90,scale=0.3]{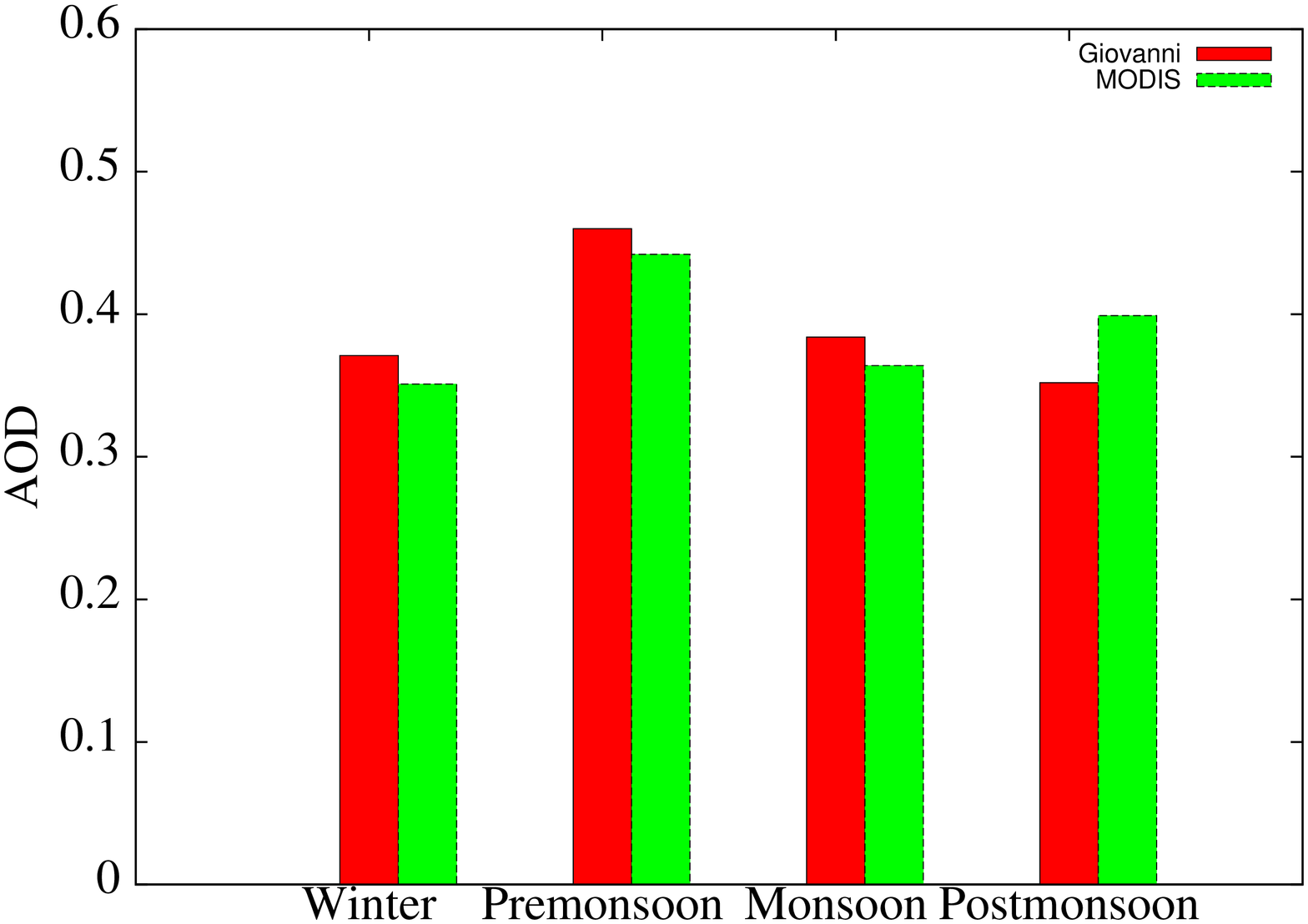}}

\caption{\label{fig:Comparison-modis&Gio}Comparison of seasonal AOD data retrieved
from MODIS level 2 data set and Giovanni level 3 data set. MODIS data
retrieved for 1 x 1 degree square area around city center whereas Giovanni
data retrieved for nearest 1 x 1 degree square grid.}

\end{figure}

\begin{figure}
\subfloat[]{\includegraphics[angle=-90,scale=0.3]{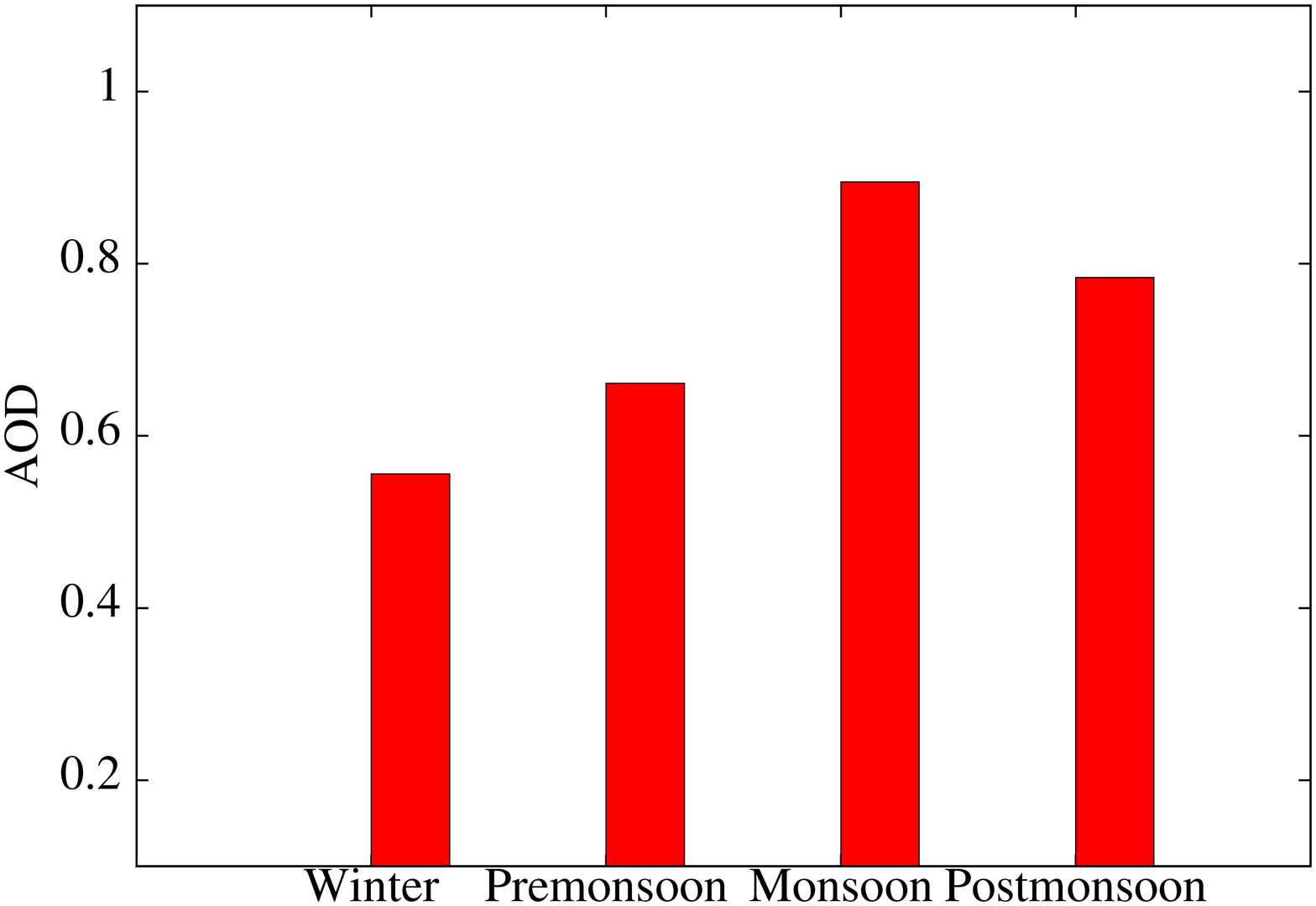}

}\subfloat[]{\includegraphics[angle=-90,scale=0.3]{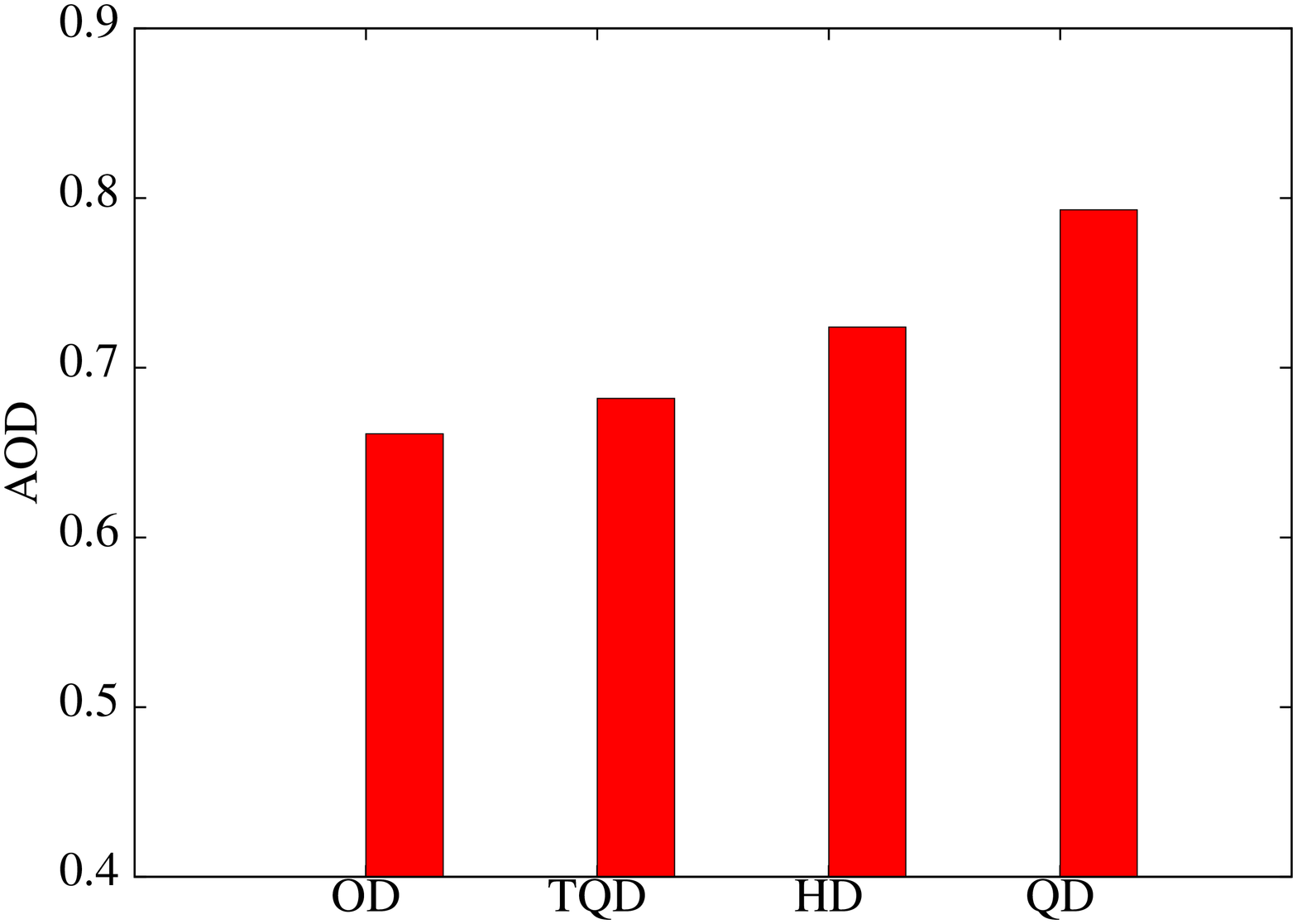}}

\caption{\label{fig-Delhi-AOD-season-spatial}(a)Delhi AOD season wise (OD
 spatial resolution ) (b) Averaged AOD for a particular season
(Premonsoon) from OD to QD over Delhi}
\end{figure}

\begin{figure}
\subfloat[Winter]{\includegraphics[angle=-90,scale=0.3]{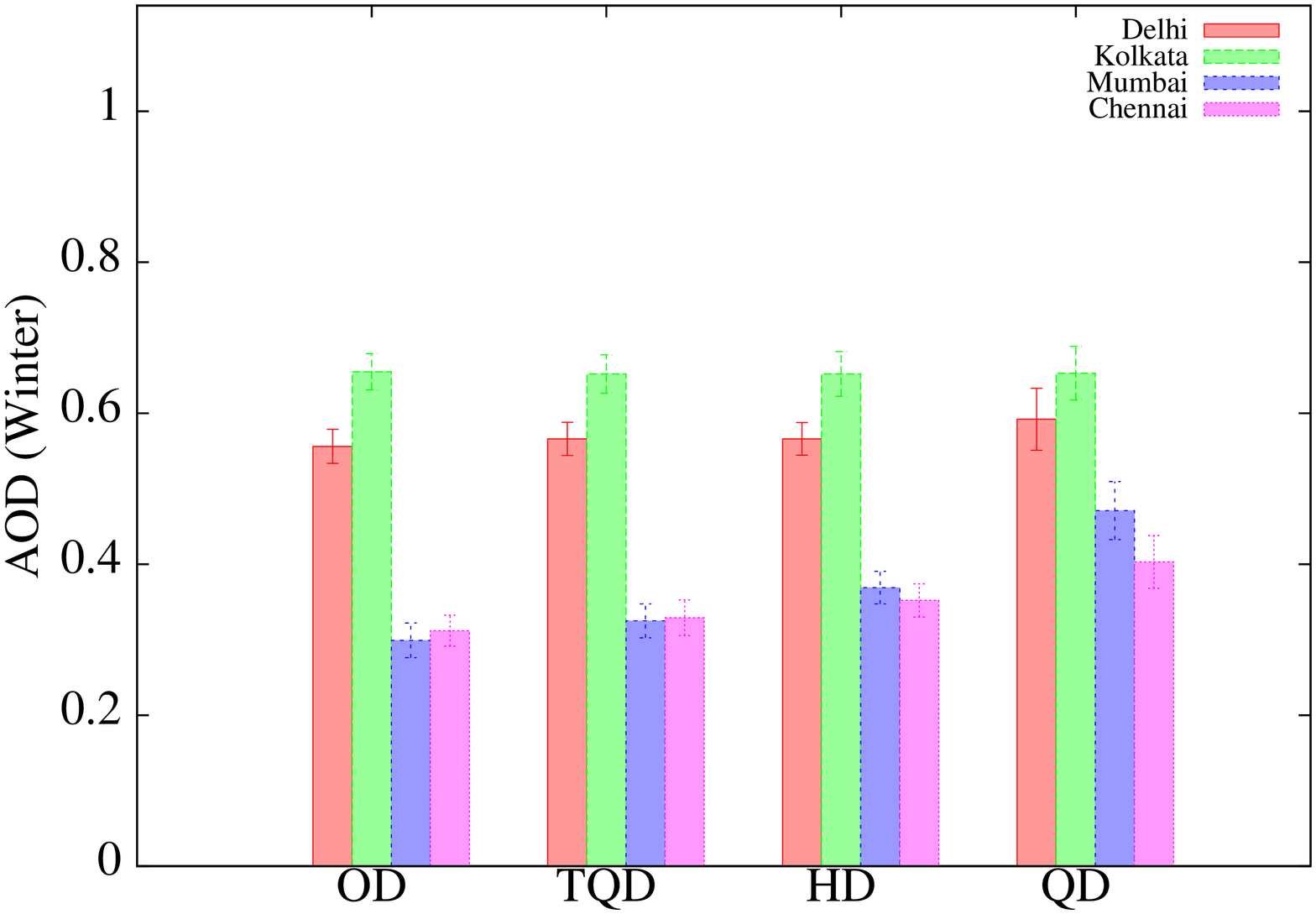}

}\subfloat[Premonsoon]{\includegraphics[angle=-90,scale=0.3]{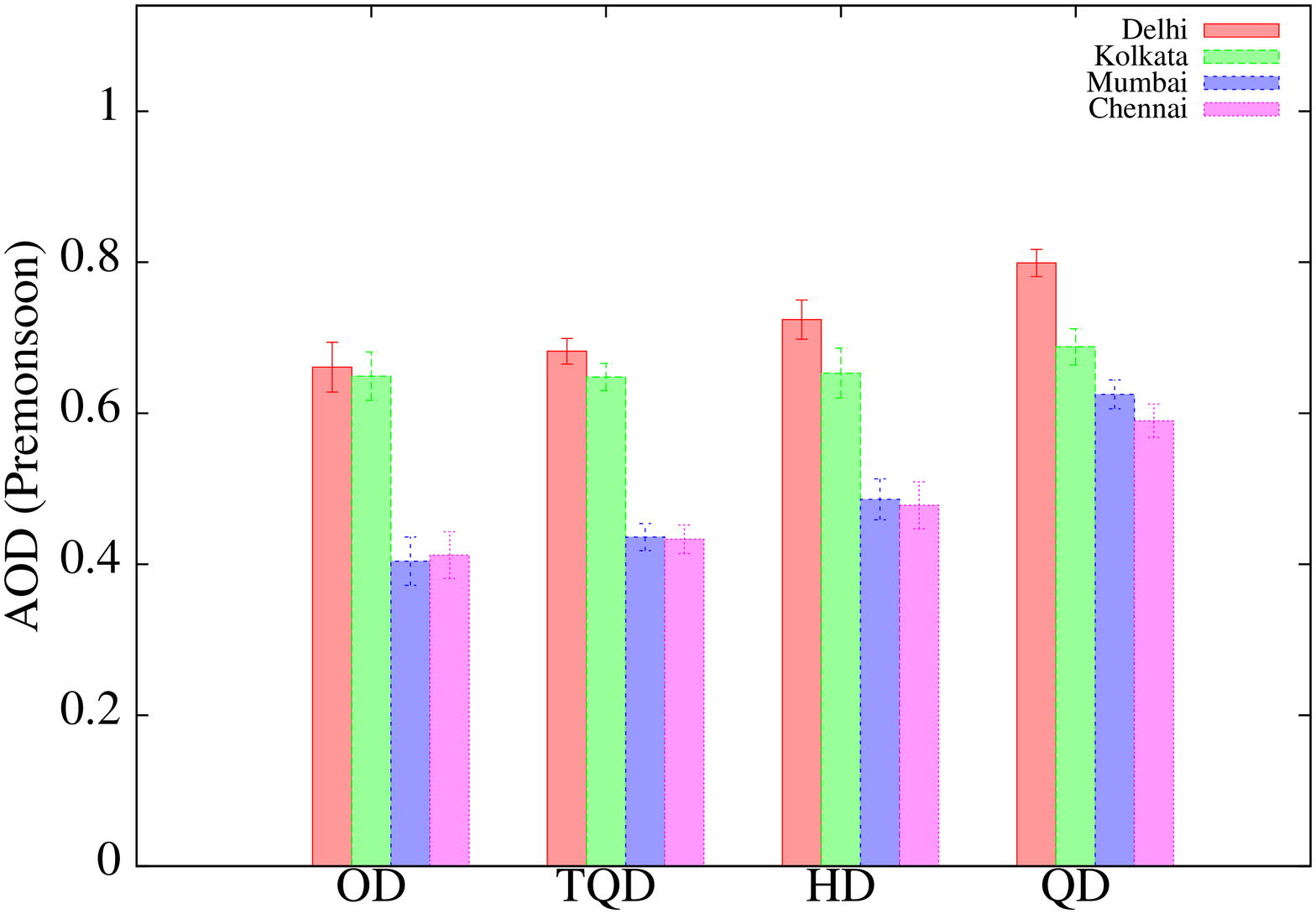}}

\subfloat[Monsoon]{\includegraphics[angle=-90,scale=0.3]{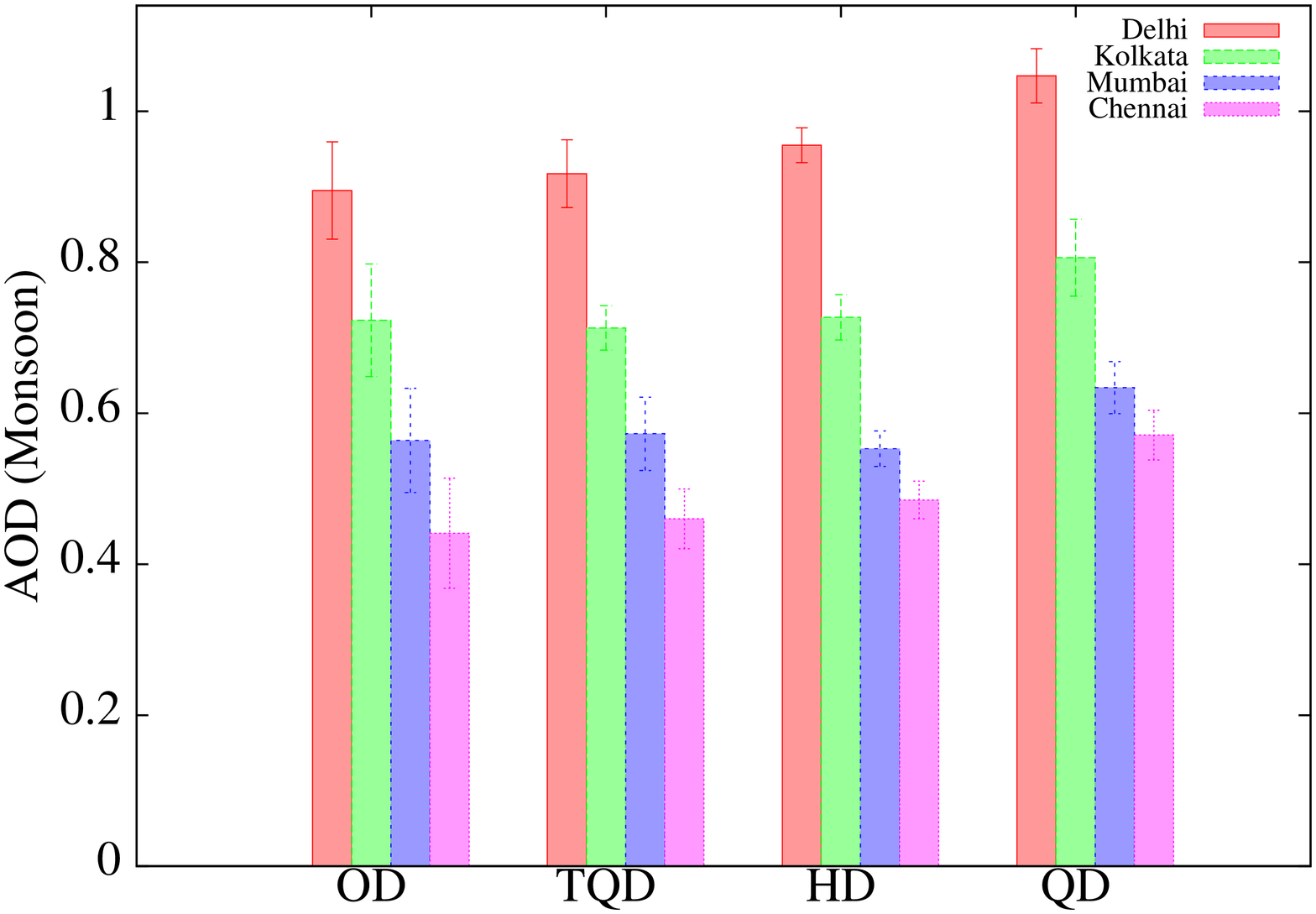}

}\subfloat[Postmonsoon]{\includegraphics[angle=-90,scale=0.3]{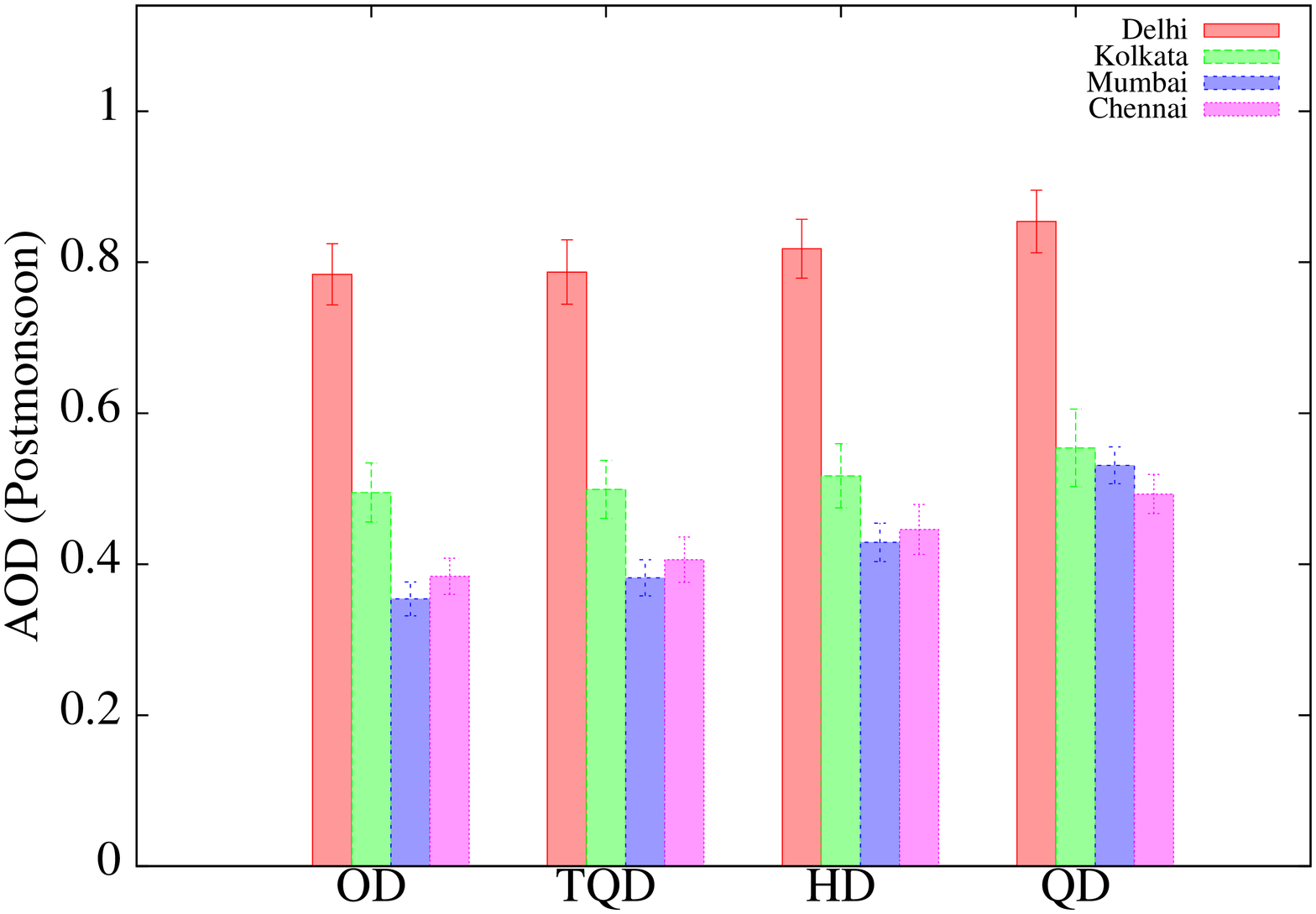}}

\caption{\label{fig:Seasonal-avergae-AOD}Seasonal average value of AOD for
four megacities in four spatial grids (error bars represent one standard deviation).}
\end{figure}

\begin{figure}

\subfloat[OD]{\includegraphics[angle=-90,scale=0.3]{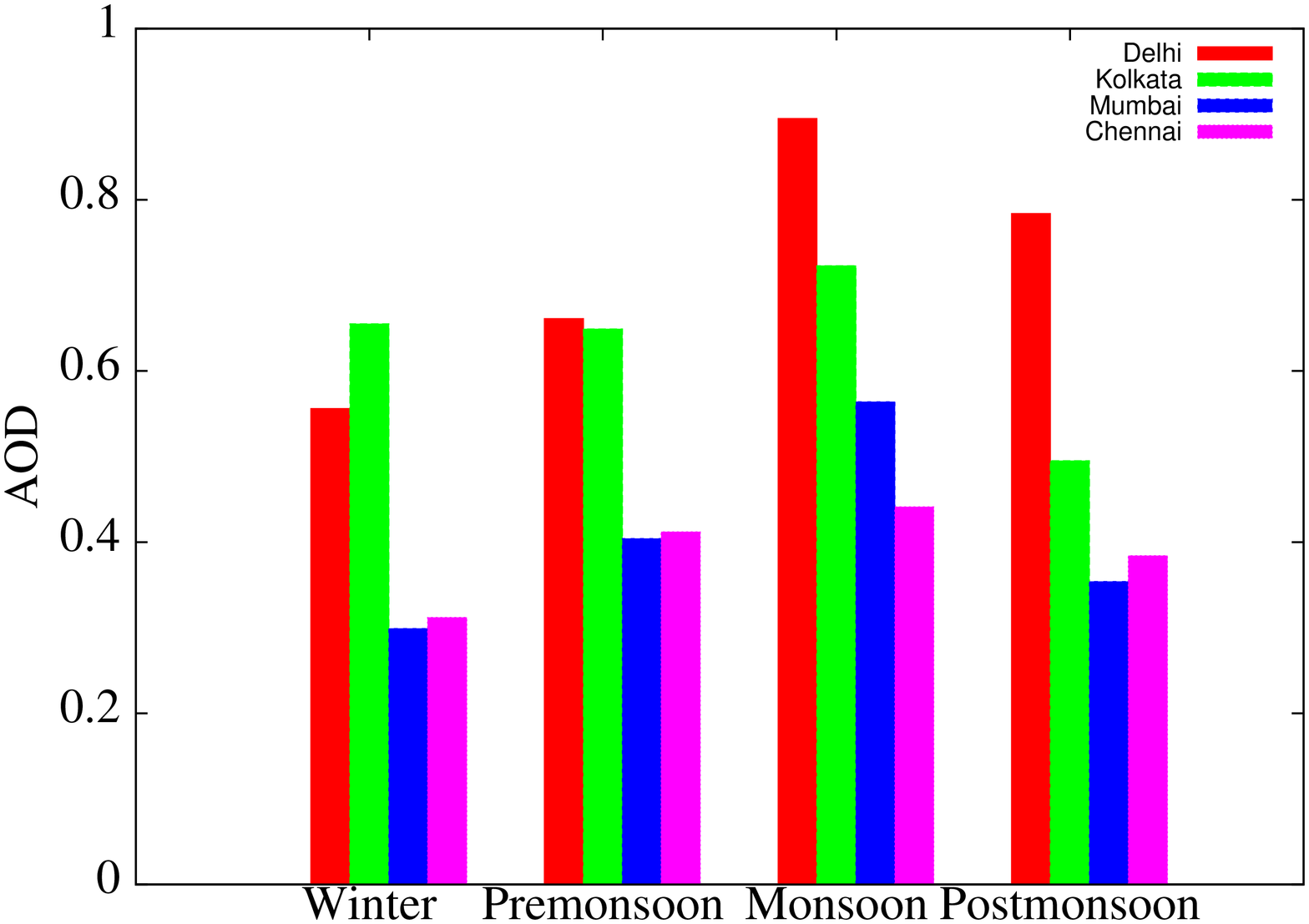}

}\subfloat[HD]{\includegraphics[angle=-90,scale=0.3]{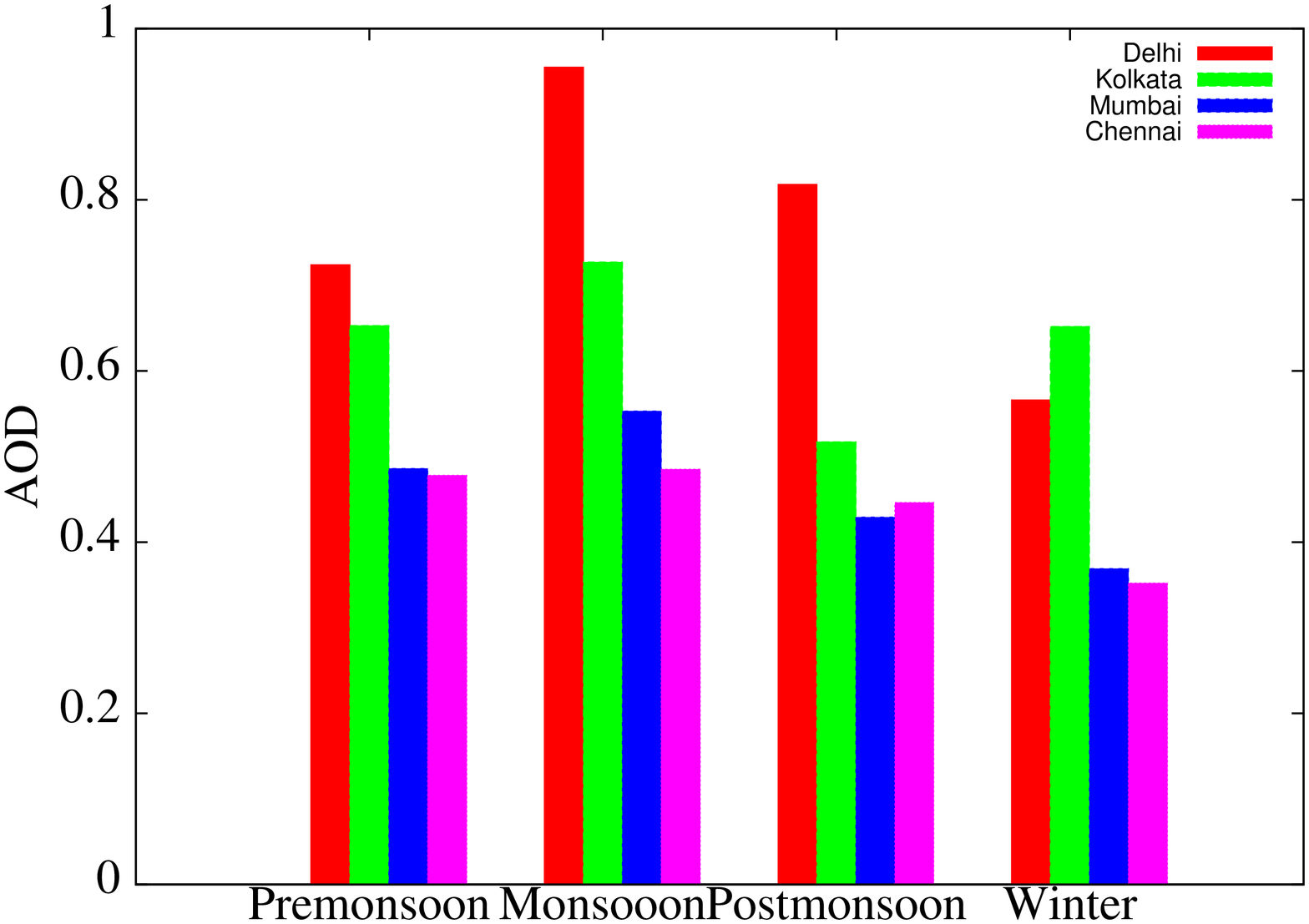}}

\caption{\label{fig:AOD-All-Cities}Comparison of seasonal average of AOD in
four megacities }

\end{figure}

\begin{figure}
\subfloat[AOD]{\includegraphics[angle=-90,scale=0.3]{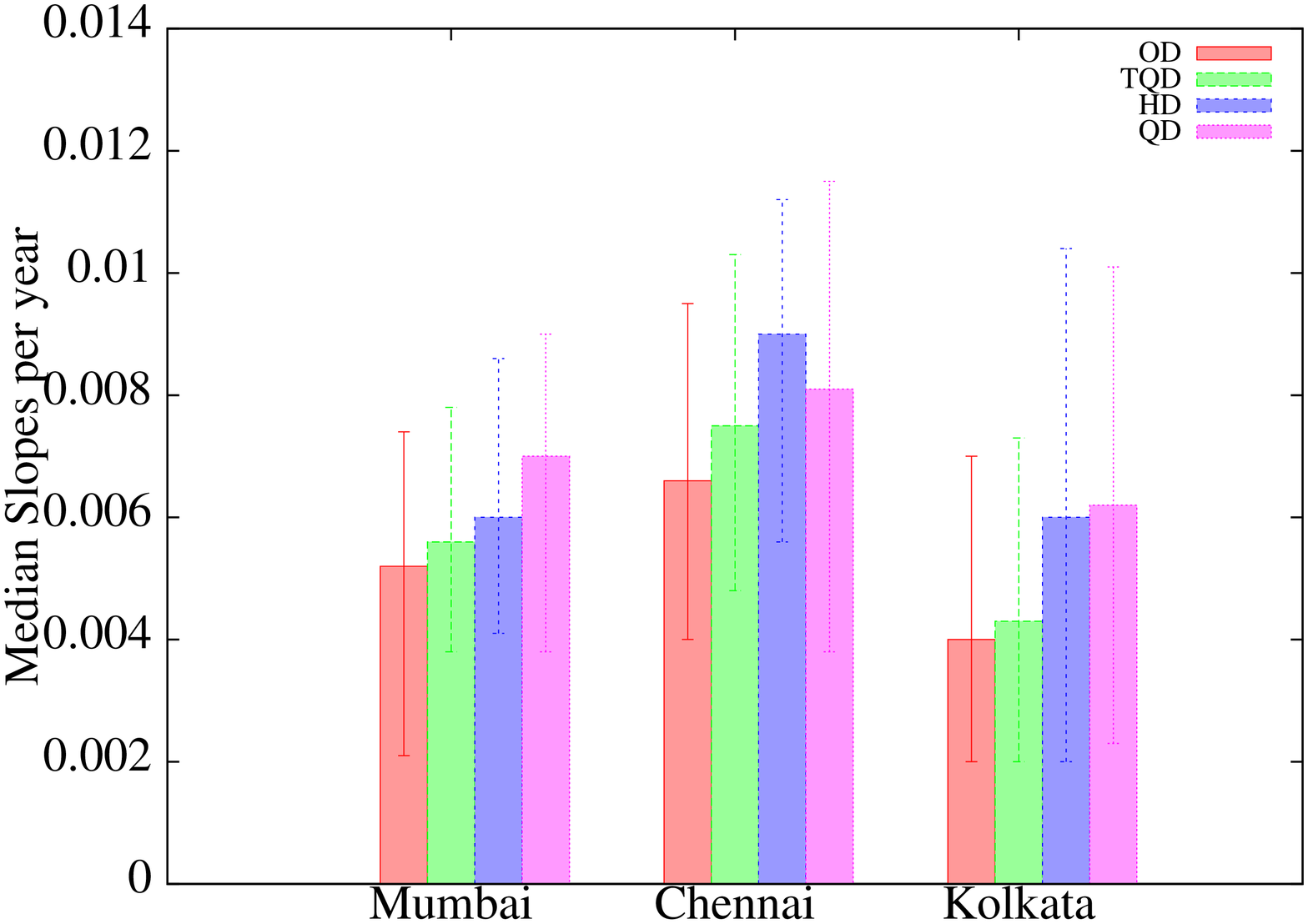}}\subfloat[\AA ngstr\"om Exponent($\alpha$) (0.47 - 0.66 $\mu m$ )]{\includegraphics[angle=-90,scale=0.3]{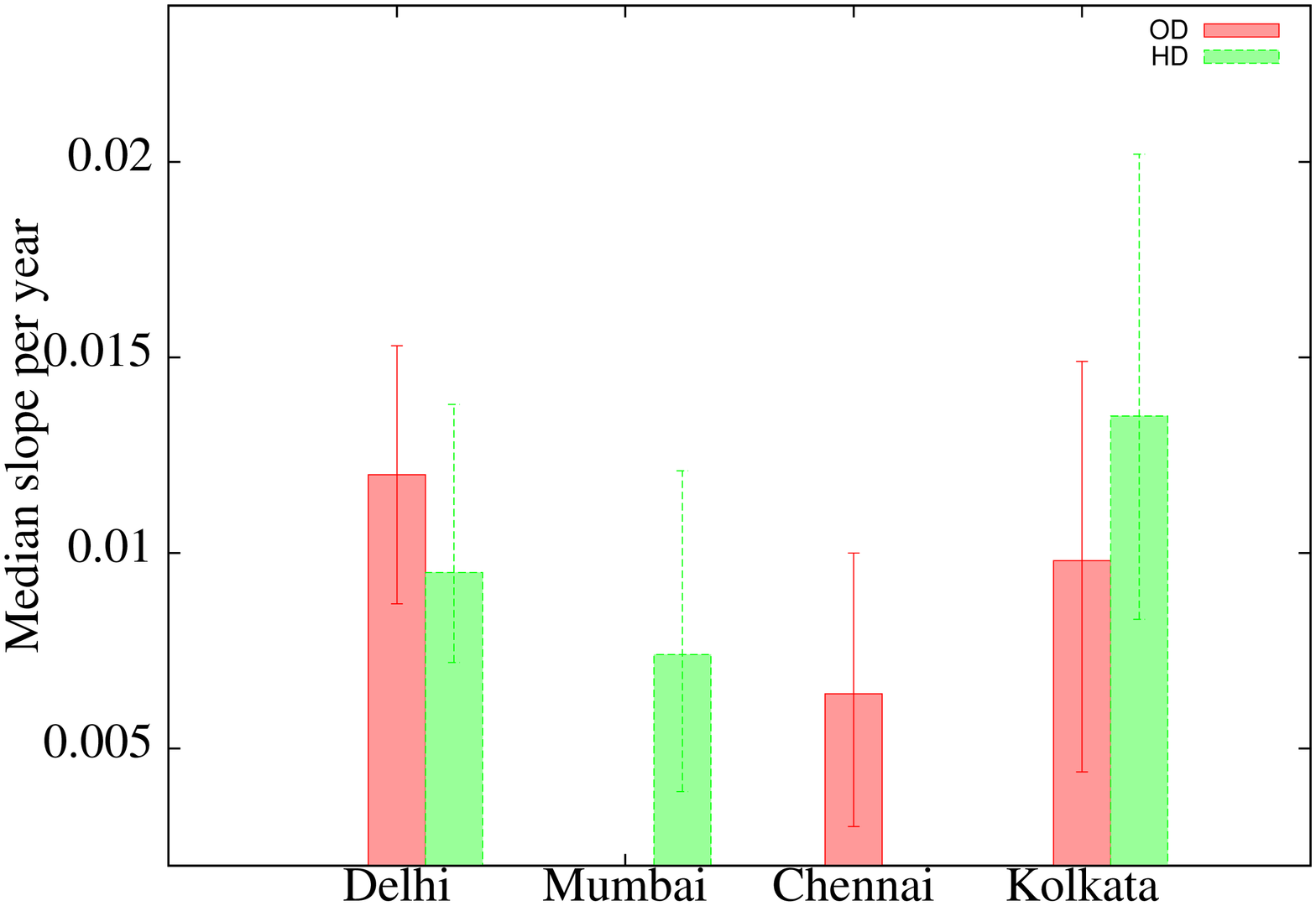}

}

\caption{\label{fig:Overall-Median-Slopes} Overall Median Slopes of trend
for A. AOD and B.  \AA ngstr\"om exponent($\alpha$) of megacities
in various  spatial resolutions (error bars represents lower and
upper limits of 80 \% confidence interval and have been calculated
using formulae \ref{M1} and \ref{M2}. )}
\end{figure}

Figure \ref{fig:Comparison-modis&Gio} compares AOD data derived from
MODIS level 2 and Giovanni level 3 for four megacities. MODIS data
derived for 1 x 1 degree square area around the city center whereas
Giovanni data are presented for nearest 1 x 1 degree grid. Figure
\ref{fig-Delhi-AOD-season-spatial}(a) shows season-wise variation
of AOD over Delhi. AOD data of last 13 years (2000-2012), averaged
season wise, has been presented. This figure clearly shows seasonal
variation. Figure \ref{fig-Delhi-AOD-season-spatial}(b) shows variation
of averaged AOD for a particular season (premonsoon here) against
different grid sizes for Delhi.
 Averaged AOD increases monotonically as we approach city center from
outskirts. Grid size for both these figure is 1 x 1 degree square.
In figure \ref{fig:Seasonal-avergae-AOD} and \ref{fig:AOD-All-Cities}
seasonal average value of AOD for four megacities have been presented
in four spatial resolutions.
 Delhi and Kolkata clearly registered higher values than Mumbai and
Chennai. However for winter seasons Kolkata registered higher values
even than Delhi. In figure \ref{fig:Overall-Median-Slopes} statistically
significant (at p=0.05 level) overall median slopes of trend derived
by Sen's method is presented for 3 megacities. Since for Delhi there
is no 'statistically significant' trend (increasing/decreasing), slopes
for Delhi have not been presented here. However from Table \ref{tab:AOD-Seasonwise}
it is evident that for seasons premonsoon and Winter Delhi showed
prominent increasing trend. Different colours shows slopes at different
grid sizes. Almost for all the cities median slopes of overall trends
increases as we approach city centres. Table \ref{tab:AOD-Seasonwise}
clearly shows for OD and TQD grid sizes almost all the cities have
increasing trend for AOD in case of winter and postmonsoon seasons.
For Chennai we find there is an increasing trend of AOD for all the
seasons in all grid sizes up to HD. Interestingly no city registered
increasing trend for AOD in any spatial grids in the season of monsoon.


\begin{table}
\label{tab:AOD-Seasonwise}\caption{Seasonal Median Slopes ( derived by Sen's method ) of aerosol optical
depth ( AOD) over four megacities ) for last 13 years (at p=0.05 level)}

\begin{tabular}{|c|c|c|c|c|c|}
\hline 
 (OD) & Winter & Premonsoon  & Monsoon  & Postmonsoon  & Overall \tabularnewline
\hline 
Delhi & $\uparrow$ 0.0087 & No & No & $\uparrow$ 0.0015 & NO\tabularnewline
\hline 
Mumbai & $\uparrow$ 0.006 & $\uparrow$ 0.0047 & No & $\uparrow$ 0.0063 & $\uparrow$ 0.0052\tabularnewline
\hline 
Chennai & $\uparrow$ 0.0081 & $\uparrow$ 0.0083 & No  & $\uparrow$ 0.0116 & $\uparrow$ 0.0066\tabularnewline
\hline 
Kolkata & $\uparrow$ 0.0123 & No  & No & $\uparrow$ 0.0085 & $\uparrow$ 0.004\tabularnewline
\hline 
\end{tabular}\\ \\

\begin{tabular}{|c|c|c|c|c|c|}
\hline 
TQD & Winter & Premonsoon & Monsoon & postmonsoon & Overall\tabularnewline
\hline 
Delhi & $\uparrow$ 0.0086 & No & No & $\uparrow$ 0.0107 & No\tabularnewline
\hline 
Mumbai & No & No & No & $\uparrow$ 0.0069 & $\uparrow$ 0.0056\tabularnewline
\hline 
Chennai & $\uparrow$ 0.0085 & $\uparrow$ 0.0067 & No & $\uparrow$ 0.0135 & $\uparrow$ 0.0075\tabularnewline
\hline 
Kolkata & $\uparrow$ 0.0115 & No & No & $\uparrow$ 0.0059 & $\uparrow$ 0.0043\tabularnewline
\hline 
\end{tabular}\\ \\

\begin{tabular}{|c|c|c|c|c|c|}
\hline 
(HD) & Winter & Premonsoon & Monsoon & Postmonsoon & Overall\tabularnewline
\hline 
Delhi & No & No & No & No & No\tabularnewline
\hline 
Mumbai & $\uparrow$ 0.008 & $\uparrow$ 0.006 & No & $\uparrow$ 0.008 & $\uparrow$ 0.006\tabularnewline
\hline 
Chennai & $\uparrow$ 0.0083 & $\uparrow$ 0.0108 & No & $\uparrow$ 0.0149 & $\uparrow$ 0.0082\tabularnewline
\hline 
Kolkata & No & No & No & No & $\uparrow$ 0.006\tabularnewline
\hline 
 &  &  &  &  & \tabularnewline
\hline 
\end{tabular}\\ \\ 

\begin{tabular}{|c|c|c|c|c|c|}
\hline 
QD & Winter & Premonsoon & Monsoon & Postmonsoon & Overall\tabularnewline
\hline 
Delhi & $\uparrow$ 0.0096 & No & No & No & No\tabularnewline
\hline 
Mumbai & No & No & -------- & $\uparrow$ 0.0081 & $\uparrow$ 0.007\tabularnewline
\hline 
Chennai & No & $\uparrow$ 0.0121 & No & No & $\uparrow$ 0.0081\tabularnewline
\hline 
Kolkata & No & No & No & No & $\uparrow$ 0.0062\tabularnewline
\hline 
Chennai & No & $\uparrow$ 0.012 & No & No & $\uparrow$ 0.008\tabularnewline
\hline 
\end{tabular}
\end{table}

\begin{figure}
\subfloat[OD]{\includegraphics[clip,angle=-90,scale=0.35]{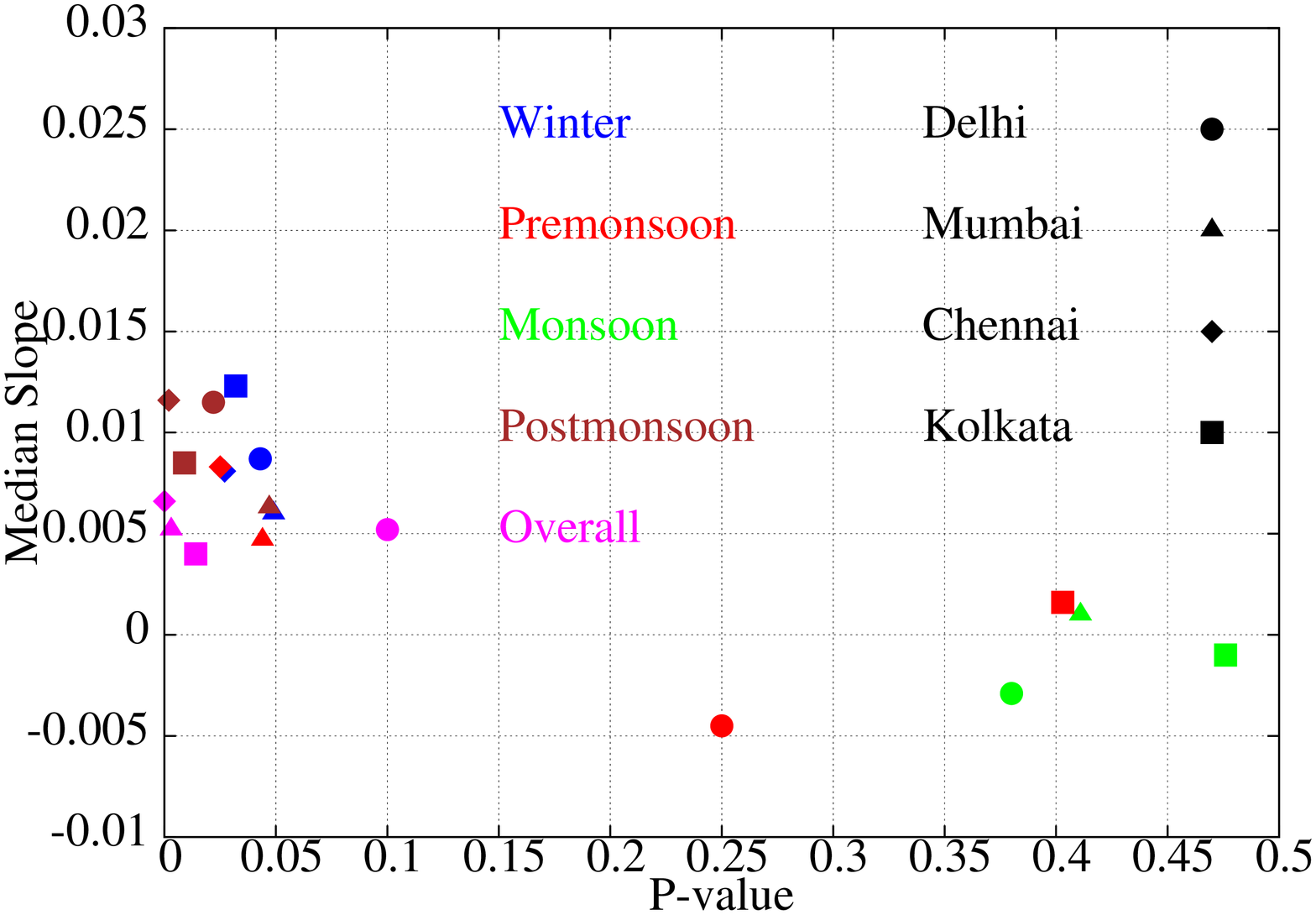}}\subfloat[TQD]{\includegraphics[clip,angle=-90,scale=0.35]{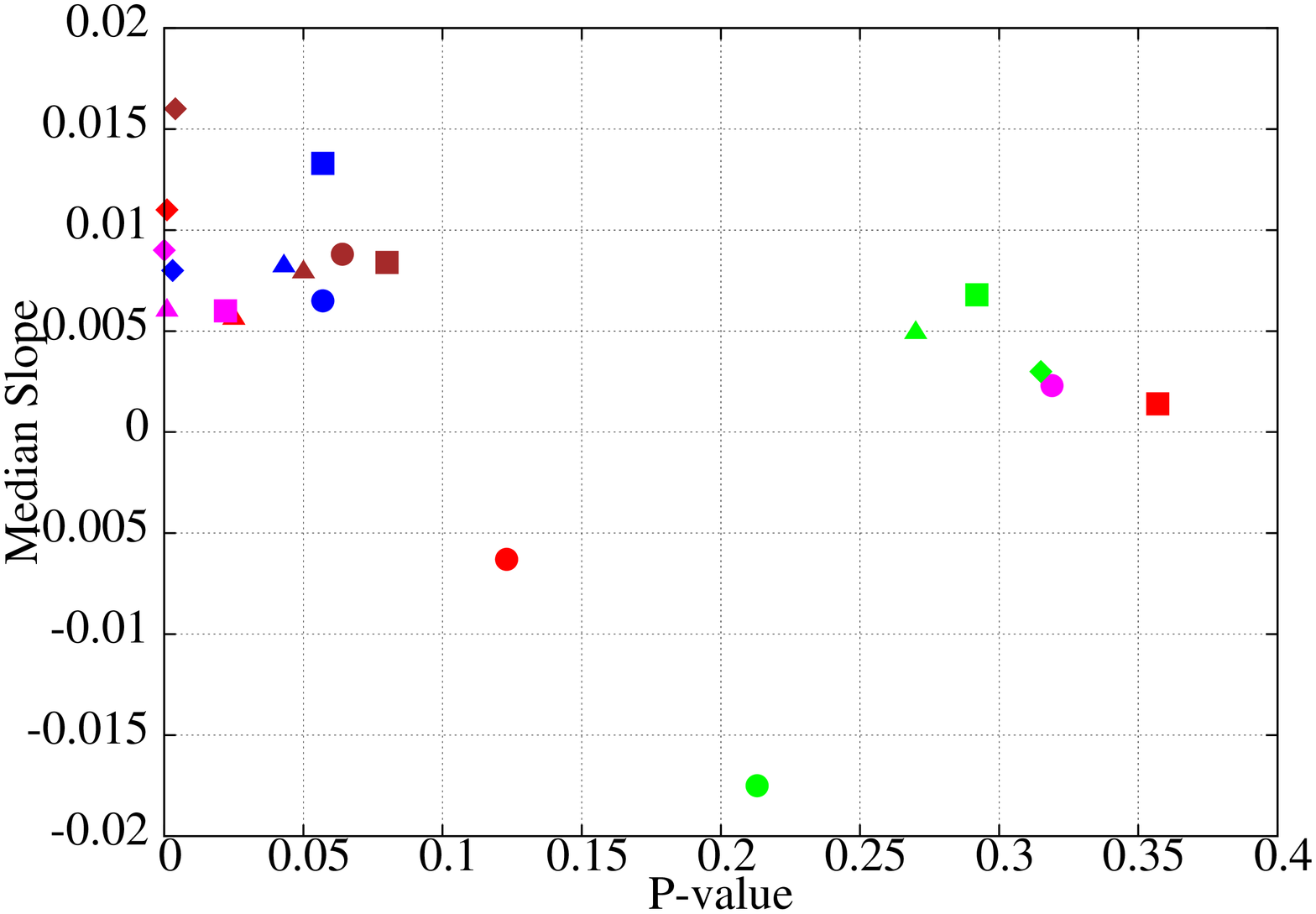}}

\subfloat[HD]{\includegraphics[clip,angle=-90,scale=0.35]{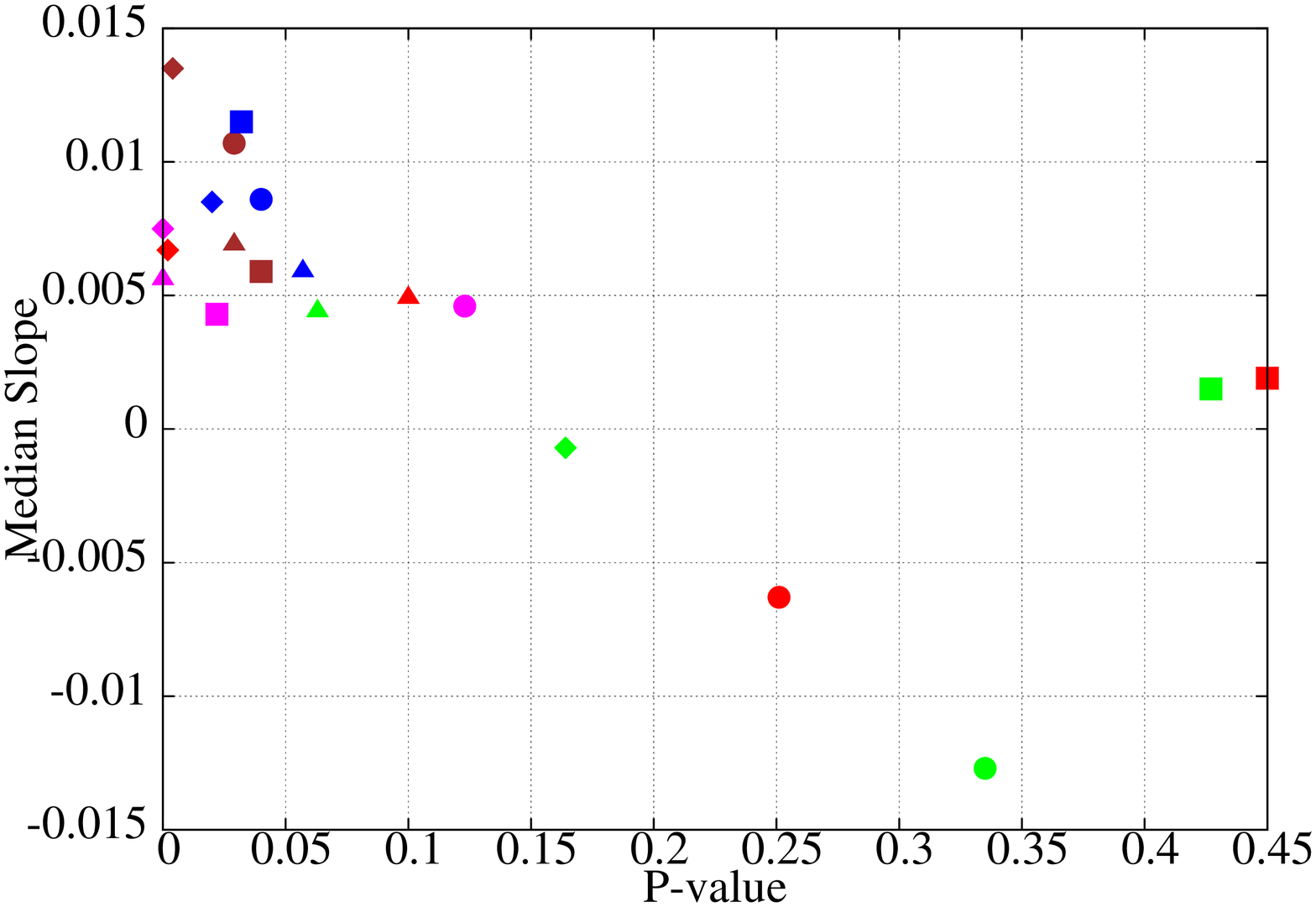}}\subfloat[QD]{\includegraphics[clip,angle=-90,scale=0.35]{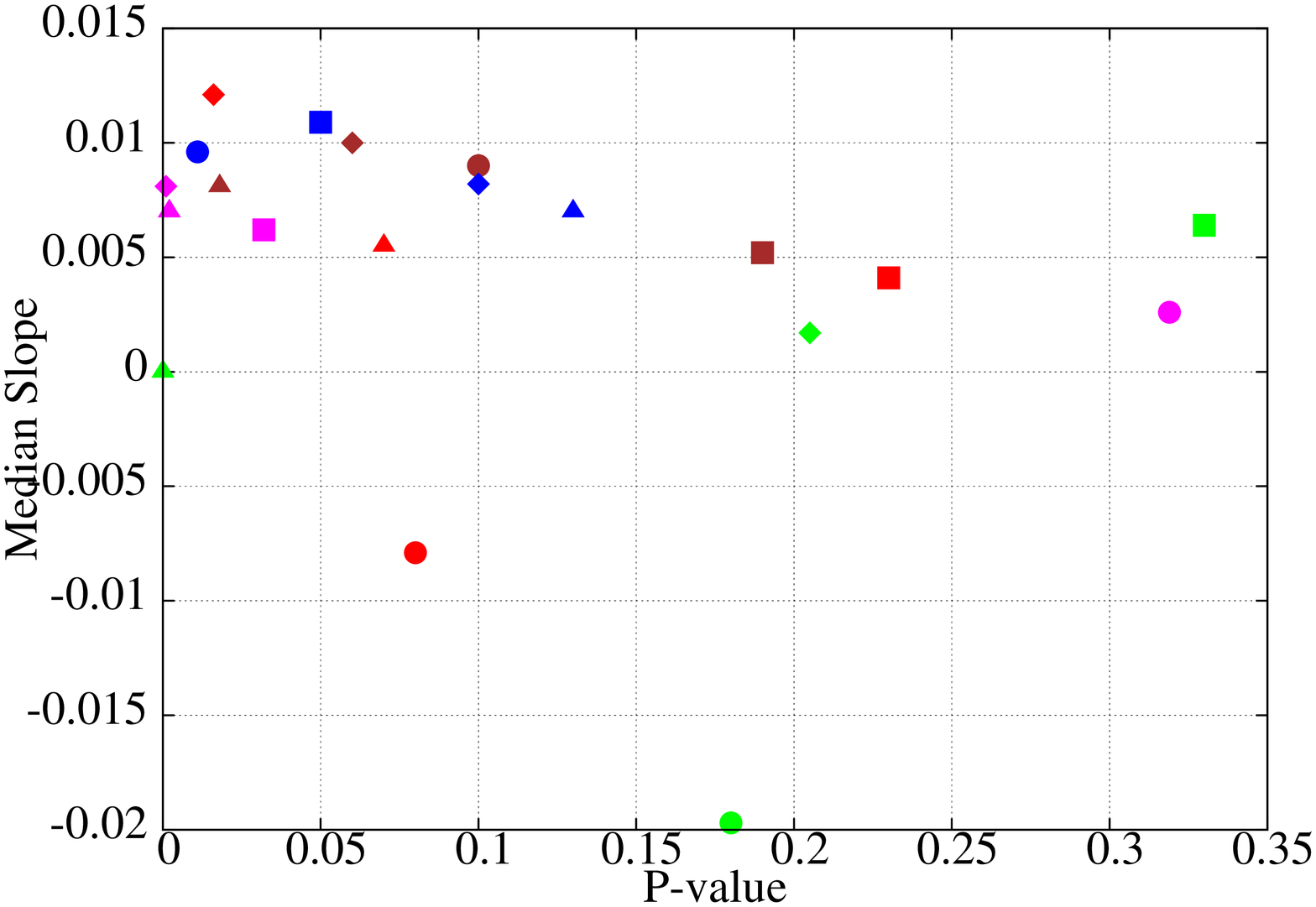}}

\caption{\label{fig:P-value-vs-Median-AOD}P-value vs Median slope of AOD trend
for four megacities in four spatial resolution}
\end{figure}

In figure \ref{fig:P-value-vs-Median-AOD} median slopes have been
presented against p-values for all four seasons and for all four megacities
including overall Kendall slope and in four grid sizes. In these figures
different colours used to present different seasons  whereas symbols
represents different cities. For  \AA ngstr\"om exponent we have
considered two spatial resolutions OD and HD. Table \ref{tab:overall med slopes angstrom}
and figure \ref{fig:Overall-Median-Slopes} (b) shows median slopes
of trend for varying spatial resolutions in case of  \AA ngsr\"om
exponent.

\begin{table}
\caption{\AA ngstr\"om exponent }

\begin{tabular}{|c|c|c|c|c|c|}
\hline 
OD & Winter & Premonsoon & Monsoon & Postmonsoon & Overall\tabularnewline
\hline 
Delhi & 0.0265 $\uparrow$ & 0.0025 $\uparrow$ & 0.0122 $\uparrow$ & 0.0254 $\uparrow$ & 0.012 $\uparrow$\tabularnewline
\hline 
Mumbai & 0.0185 $\uparrow$ & No & -0.0156 $\downarrow$ & 0.0169 $\uparrow$ & No\tabularnewline
\hline 
Chennai & No & No  & No  & No  & 0.0064 $\uparrow$\tabularnewline
\hline 
Kolkata & 0.0253 $\uparrow$ & 0.0083 $\uparrow$  & No & No & 0.0098 $\uparrow$\tabularnewline
\hline 
\end{tabular}\\  \\

\begin{tabular}{|c|c|c|c|c|c|}
\hline 
HD & Winter & Premonsoon & Monsoon & Postmonsoon & Overall\tabularnewline
\hline 
Delhi & 0.0216 $\uparrow$ & 0.0031 $\uparrow$ & 0.0088 $\uparrow$  & 0.0229 $\uparrow$ & 0.0095 $\uparrow$\tabularnewline
\hline 
Mumbai & No & No & No & 0.0194 $\uparrow$ & 0.0074 $\uparrow$\tabularnewline
\hline 
Chennai & No & No & No & No & No\tabularnewline
\hline 
Kolkata & 0.0251 $\uparrow$ & 0.0113 $\uparrow$ & No & 0.0241 $\uparrow$ & 0.0135 $\uparrow$\tabularnewline
\hline 
\end{tabular}
\end{table}

\begin{table}
\caption{Comparison of Median slopes (Overall) of aerosol optical depth ( AOD)
for 13 years for four megacities in four spatial resolutions (at p=0.05
level)}

\centering{}%
\begin{tabular}{|c|c|c|c|c|}
\hline 
 & OD & TQD & HD & QD\tabularnewline
\hline 
Delhi & No & No & No & No\tabularnewline
\hline 
Mumbai & $\uparrow$ 0.0052 & $\uparrow$ 0.0056 & $\uparrow$ 0.006 & $\uparrow$ 0.007\tabularnewline
\hline 
Chennai & $\uparrow$ 0.0066 & $\uparrow$ 0.0075 & $\uparrow$ 0.006 & $\uparrow$ 0.00862\tabularnewline
\hline 
Kolkata & $\uparrow$ 0.004 & $\uparrow$ 0.0043 & $\uparrow$ 0.006 & $\uparrow$ 0.0062\tabularnewline
\hline 
\end{tabular}
\end{table}


The study clearly shows AOD is increasing in three of the four Indian
megacities even when seasonal effects are considered. Only Delhi registered
no significant overall increase corroborating earlier studies but
one should remember AOD over Delhi is already very high compared to
other cities. However Delhi registered significant increasing trend
in winter and postmonsoon seasons in some spatial grids{[}Table \ref{tab:AOD-Seasonwise},
fig. \ref{fig:P-value-vs-Median-AOD}{]}. City centres not only are
already high in AOD, rate of increase is also higher in city centres
compared to outskirts. In case of  \AA ngstr\"om exponent the scenario
is more or less the same. Since higher  \AA ngstr\"om exponent means
lower particle size, increase in this exponent points to more hazards
to public health because smaller particles are known to cause more
severe problems\cite{pope-2,pope-1}. The case of Delhi should be
considered here specially, Delhi has not registered any significant
growth in AOD but it has shown significant increasing trend in case
of  \AA ngstr\"om exponent. Although ground based measurements are
best to reach any concrete conclusion, satellite sensed data averaged
by seasons and compared for long time undoubtedly gives an overview
how and at which rate our cities are getting polluted as time goes
and hopefully this study will come handy to policymakers and researcher
for more detailed studies. 
\begin{table}
\begin{centering}
\caption{\label{tab:overall med slopes angstrom}Comparison of Median slopes
(Overall)  \AA ngstr\"om exponent for 13 years for four megacities
in two spatial resolutions (at p=0.05 level)}

\par\end{centering}

\centering{}%
\begin{tabular}{|c|c|c|}
\hline 
 & OD & HD\tabularnewline
\hline 
Delhi & 0.012 $\uparrow$ & 0.0095 $\uparrow$\tabularnewline
\hline 
Mumbai & NO & 0.0074 $\uparrow$ \tabularnewline
\hline 
Chennai & 0.0064 $\uparrow$ & No\tabularnewline
\hline 
Kolkata & 0.0098 $\uparrow$ & 0.0135 $\uparrow$\tabularnewline
\hline 
\end{tabular}
\end{table}

\begin{figure}

\subfloat[Grid size OD]{\includegraphics[clip,angle=-90,scale=0.35]{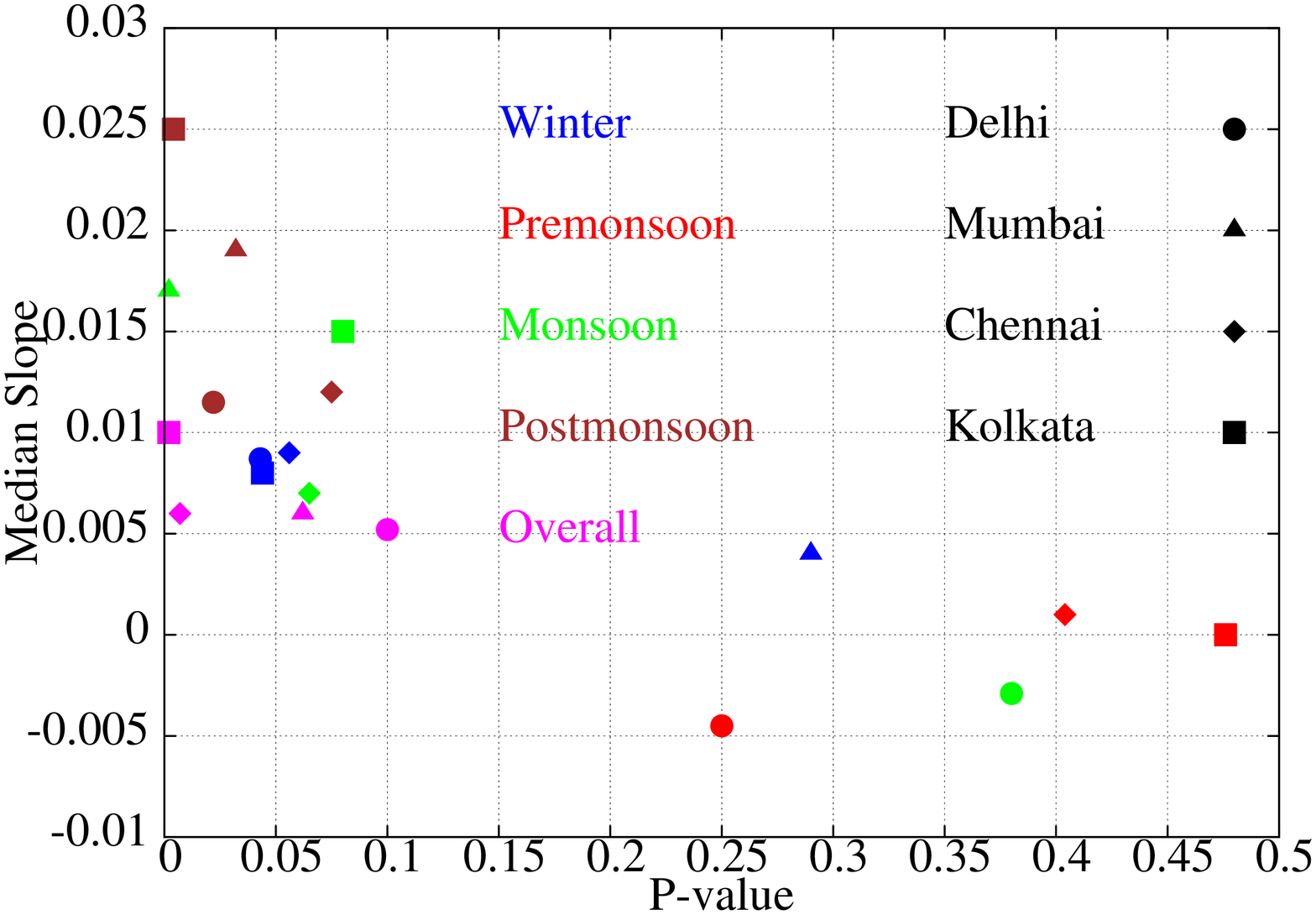}

}\subfloat[Grid size HD]{\includegraphics[clip,angle=-90,scale=0.35]{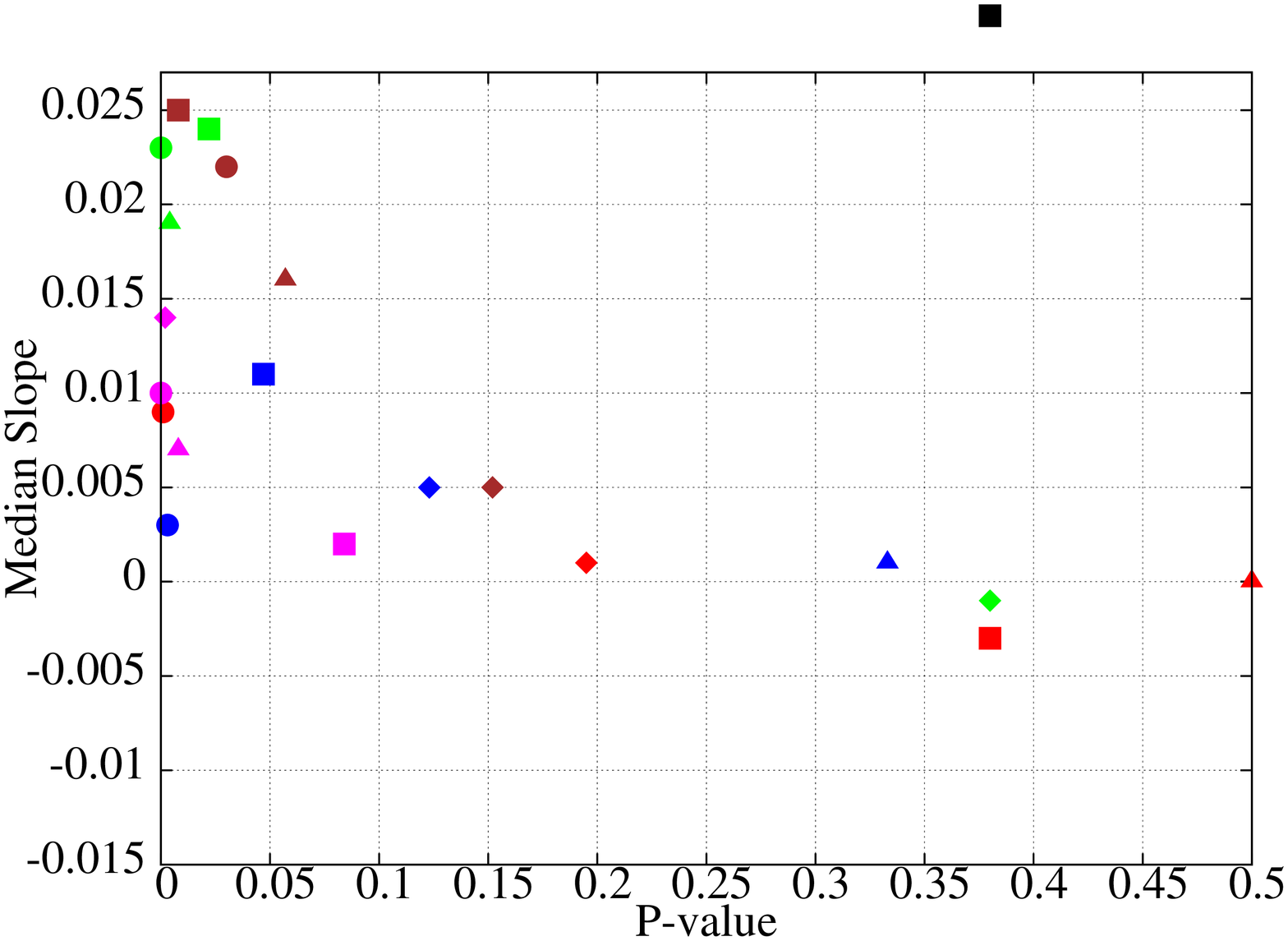}}

\caption{Median slopes of trend vs p-value of  \AA ngstr\"om exponent($\alpha$)
for four megacities in two spatial resolution}
\end{figure}


Acknowledgements:

We thank the entire MODIS science team for providing us data used
in this study.

\end{document}